\documentclass[11pt]{article}\textwidth 6.5in\textheight 9in
\usepackage{amssymb}\usepackage[colorlinks]{hyperref}\usepackage{color}
	\usepackage{stmaryrd}\usepackage{mathrsfs}
	\usepackage{graphicx}
	\usepackage{amsmath}
\usepackage{caption}

\begin{document}
\setlength{\captionmargin}{27pt}
\newcommand\hreff[1]{\href {http://#1} {\small http://#1}}
\newcommand\trm[1]{{\bf\em #1}} \newcommand\emm[1]{{\ensuremath{#1}}}
\newcommand\prf{\paragraph{Proof.}}\newcommand\qed{\hfill\emm\blacksquare}

\setcounter{tocdepth}{3} 

\newtheorem{thr}{Theorem} 
\newtheorem{lmm}{Lemma}
\newtheorem{cor}{Corollary}
\newtheorem{con}{Conjecture} 
\newtheorem{prp}{Proposition}

\newtheorem{blk}{Block}
\newtheorem{dff}{Definition}
\newtheorem{asm}{Assumption}
\newtheorem{rmk}{Remark}
\newtheorem{clm}{Claim}
\newtheorem{example}{Example}

\newcommand{\ab}{a\!b}
\newcommand{\yx}{y\!x}
\newcommand{\yux}{y\!\underline{x}}

\newcommand\floor[1]{{\lfloor#1\rfloor}}\newcommand\ceil[1]{{\lceil#1\rceil}}

\newcommand{\lea}{<^+}
\newcommand{\gea}{>^+}
\newcommand{\eqa}{=^+}

\newcommand{\lel}{<^{\log}}
\newcommand{\gel}{>^{\log}}
\newcommand{\eql}{=^{\log}}

\newcommand{\lem}{\stackrel{\ast}{<}}
\newcommand{\gem}{\stackrel{\ast}{>}}
\newcommand{\eqm}{\stackrel{\ast}{=}}

\newcommand\edf{{\,\stackrel{\mbox{\tiny def}}=\,}}
\newcommand\edl{{\,\stackrel{\mbox{\tiny def}}\leq\,}}
\newcommand\then{\Rightarrow}

\newcommand\km{{\mathbf {km}}}\renewcommand\t{{\mathbf {t}}}
\newcommand\KM{{\mathbf {KM}}}\newcommand\m{{\mathbf {m}}}
\newcommand\md{{\mathbf {m}_{\mathbf{d}}}}\newcommand\mT{{\mathbf {m}_{\mathbf{T}}}}
\newcommand\K{{\mathbf K}} \newcommand\I{{\mathbf I}}

\newcommand\II{\hat{\mathbf I}}
\newcommand\Kd{{\mathbf{Kd}}} \newcommand\KT{{\mathbf{KT}}} 
\renewcommand\d{{\mathbf d}} 
\newcommand\D{{\mathbf D}}

\newcommand\w{{\mathbf w}}
\newcommand\Ks{\Lambda} \newcommand\q{{\mathbf q}}
\newcommand\E{{\mathbf E}} \newcommand\St{{\mathbf S}}
\newcommand\M{{\mathbf M}}\newcommand\Q{{\mathbf Q}}
\newcommand\ch{{\mathcal H}} \renewcommand\l{\tau}
\newcommand\tb{{\mathbf t}} \renewcommand\L{{\mathbf L}}
\newcommand\bb{{\mathbf {bb}}}\newcommand\Km{{\mathbf {Km}}}
\renewcommand\q{{\mathbf q}}\newcommand\J{{\mathbf J}}
\newcommand\z{\mathbf{z}}

\newcommand\B{\mathbf{bb}}\newcommand\f{\mathbf{f}}
\newcommand\hd{\mathbf{0'}} \newcommand\T{{\mathbf T}}
\newcommand\R{\mathbb{R}}\renewcommand\Q{\mathbb{Q}}
\newcommand\N{\mathbb{N}}\newcommand\BT{\{0,1\}}
\newcommand\FS{\BT^*}\newcommand\IS{\BT^\infty}
\newcommand\FIS{\BT^{*\infty}}\newcommand\C{\mathcal{L}}
\renewcommand\S{\mathcal{C}}\newcommand\ST{\mathcal{S}}
\newcommand\UM{\nu_0}\newcommand\EN{\mathcal{W}}

\newcommand{\supp}{\mathrm{Supp}}

\newcommand\lenum{\lbrack\!\lbrack}
\newcommand\renum{\rbrack\!\rbrack}

\renewcommand\qed{\hfill\emm\square}

\title{\vspace*{-3pc} On the Existence of Anomalies}

\author {Samuel Epstein\footnote{JP Theory Group. samepst@jptheorygroup.org}}

\maketitle
\begin{abstract}
The Independence Postulate (IP) is a finitary Church-Turing Thesis, saying mathematical sequences are independent from physical ones. IP implies the existence of anomalies. 
\end{abstract}
\section{Introduction}
An anomaly is a measurement that deviates so much from other observations as to arouse suspicion that it was generated by a different mechanism. In previous work, \cite{Epstein21}, it has been proved that algorithmic sampling methods have to produce anomalies. However some sampling methods are too complex to be considered algorithmic. One example is your local weather forecast. Using the Independence Postulate, which is a finitary Church-Turing thesis, this open issue is addressed. Outliers must occur in the physical world.
\section{Related Work}
The study of Kolmogorov complexity originated from the work of~\cite{Kolmogorov65}. The canonical self-delimiting form of Kolmogorov complexity was introduced in~\cite{ZvonkinLe70} and~\cite{Chaitin75}. The Independence Postulate was introduced in \cite{Levin84,Levin13} and intially used to show the impossibility of consistent completions of Peano Arithmetic, and later generalized in \cite{BienvenuPo16}. This article uses notions of the mutual information with the halting sequence, and more information about this term can be found in \cite{Epstein21,VereshchaginVi02}. 
\section{Conventions}

The function $\K(x|y)$ is the conditional prefix Kolmogorov complexity. The mutual information between two strings $x,y\in\FS$, is $\I(x:y)=\K(x)+\K(y)-\K(x,y)$. For probability $p$ over $\N$, randomness deficiency is $\d(a|p,b)=\floor{-\log p(a)}-\K(a|\langle p\rangle, b)$ and measures the extent of the refutation against the hypothesis $p$ given the result $a$ \cite{Gacs21}. $\d(a|p)=\d(a|p,\emptyset)$. The amount of information that the halting sequence $\ch\in\IS$ has about $a\in\FS$ is $\I(a;\ch)=\K(a)-\K(a|\ch)$. We use ${\lea} f$ to denote ${<}f{+}O(1)$ and ${\lel} f$ to denote ${<}f {+} O(\log(f{+}1))$.  Stochasticity is $\Ks(a|b)= \min\{\K(Q|b)+3\log\max\{\d(a|Q,b),1\}
:\textrm{$Q$ has finite support and a range in $\Q$}\}$.
$\Ks(a|b)$ $<\Ks(a)+O(\log\K(b))$.
The following definition is from \cite{Levin74} .
\begin{dff}[Information]For infinite sequences $\alpha,\beta\in\IS$, their mutual information is defined to be 
$\I(\alpha\,{:}\,\beta){=}$
 $\log\sum_{x,y\in\FS}2^{\I(x:y)-\K(x|\alpha)-\K(y|\beta)}$.
\end{dff}

\noindent The Independence Postulate (\textbf{IP}), \cite{Levin84,Levin13}, is an unprovable inequality on the information content shared between two sequences. \textbf{IP} is a finitary Church Turing Thesis, postulating that certain infinite and finite sequences cannot be found in nature, a.k.a. have high “physical addresses”.\\

\noindent \textbf{IP}\textit{: Let $\alpha$ be a sequence defined with an $n$-bit mathematical statement, and a sequence $\beta$ can be located in the physical world with a $k$-bit instruction set. Then $\I(\alpha:\beta)<k+n+c$ for some small absolute constant $c$.}
\section{Results}

There are many proofs in the literature that non-stochastic numbers have high mutual information with the halting sequence. One such detailed proof is in \cite{Epstein21}.
\begin{lmm}
	\label{lmm}
	$\Ks(x)\lel\I(x;\ch)$.
\end{lmm}

\begin{lmm}
\label{lmm2}
For probability $p$ over $\N$, $D{\subset}\N$, $|D|=2^s$, $s < \max_{a\in D}\d(a|p)+\Ks(D)+\K(s)+O(\log \K(s,p))$. 
\end{lmm}
\begin{prf}
We relativize the universal Turing machine to $\langle s,p\rangle$. Let $Q$ be a probability measure that realizes $\Ks(D)$, with $d=\max\{\d(D|Q),1\}$. Let $F\subseteq\N$ be a random set where each element $a\in\N$ is selected independently with probability $cd2^{-s}$, where $c\in\N$ is chosen later. $\E[p(F)]\leq cd2^{-s}$. Furthermore 
\begin{align*}
&\E[Q(\{G:|G|=2^s,G\cap F=\emptyset\})]
\leq \sum_GQ(G)(1-cd2^{-s})^{2^s}< e^{-cd}.
\end{align*}
 Thus finite $W\subset\N$ can be chosen such that $p(W)\leq 2cd2^{-s}$ and $Q(\{G:|G|=2^s,G\cap W=\emptyset\}) \leq e^{1-cd}$. $D\cap W\neq \emptyset$, otherwise, using the $Q$-test, $t(G)=e^{cd-1}$ if $(|G|=2^s, G\cap W=\emptyset)$ and $t(G)=0$ otherwise, we have
\begin{align*}
\K(D|Q,d,c) & \lea -\log Q(D)-(\log e)cd\\
(\log e)cd & \lea -\log Q(D)-\K(D|Q) + \K(d,c) \\
(\log e)cd & \lea d + \K(d,c),
\end{align*}
which is a contradiction for large enough $c$. Thus there is an $a\in D\cap W$, where 
\begin{align*}
\K(a) &\lea -\log p(a) +\log d -s +\K(d)+\K(Q)\\
s&\lea \d(a|p)+\Ks(D).
\end{align*}
Making the relativization of $\langle s,p\rangle$ explicit,
\begin{align*}
s&< -\log p(a)-\K(a|s,p)+\Ks(D|s,p)\\
s&< \max_{a\in D}\d(a|p)+\Ks(D)+\K(s)\\
&+O(\log\K(s,p)).\;\,\qed
\end{align*}
\end{prf}
Let $\tau{\in}\N^\N$ represent a series of observations. In reality, observed information is finite. But observations can be considered to be potentially infinite, and represented by never-ending sequences. Assuming $\tau$ has an infinite amount of unique numbers, $\tau(n)$ is the first $2^n$ unique numbers of $\tau$.\newpage
\begin{thr}
\label{thr:infpref}
For probability $p$ over $\N$, $\tau{\in}\N^\N$, let $s_{\tau,p}=\sup_n\left( n - 3\K(n)-\max_{a\in\tau(n)}\d(a|p)\right)$. Then $s_{\tau,p}\lel \I(\langle \tau\rangle:\ch)+O(\log \K(p))$.
\end{thr}
\begin{prf} By Lemmas \ref{lmm} and \ref{lmm2},  and the fact that $\I(x;\ch)\lea \I(\alpha:\ch)+\K(x|\alpha)$,
\begin{align*}
n&< \max_{a\in \tau(n)} \d(a|p) +\I(\tau(n);\ch)++\K(n)
+O(\log\I(\tau(n);\ch)\K(p)\K(n)),\\
n&< \max_{a\in \tau(n)} \d(a|p) +2\K(n)+\I(\langle\tau\rangle:\ch)+O(\log\I(\langle\tau\rangle:\ch)\K(p)\K(n)),\\
n&{-} {3\K(n)} {-} {\max_{a\in \tau(n)}} \d(a|p){\lel}\I(\langle\tau\rangle{:}\ch){+}O(\log\K(p)).
\qed
\end{align*}
\end{prf}
 Let $k$ be a physical address of $\tau$. $\ch$ can be described by a small mathematical statement. By Theorem \ref{thr:infpref} and \textbf{IP},
\begin{align*}
s_{\tau,p}&\lel \I(\langle\tau\rangle:\ch)+O(\log \K(p))\lel k+ c +O(\log \K(p)).
\end{align*}
It's hard to find observations with small anomalies and impossible to find observations with no anomalies.
\section{Discussion}
This article shows that sequences of observations $\tau$ will contain ever increasing anomalies $a$, i.e. high $\d(a|p)$ for some probability $p$, otherwise it can't be reached by a physical address. In this article, observations are modelled by infinite sequences of natural numbers. However, with some work, this can generalized to infinite sequences of real numbers. Further work involves associating the notion of ``outlier'' or ``anomaly'' to more general topologies, such as computable metric spaces.


\begin{thebibliography}{Cha75}

\bibitem[BP16]{BienvenuPo16}
L.~Bienvenu and C.~Porter.
\newblock Deep ${\Pi} _1^0$ classes.
\newblock {\em The Bulletin of Symbolic Logic}, 22(2):249--286, 2016.

\bibitem[Cha75]{Chaitin75}
G.~J. Chaitin.
\newblock {A Theory of Program Size Formally Identical to Information Theory}.
\newblock {\em Journal of the ACM}, 22(3):329--340, 1975.

\bibitem[Eps21]{Epstein21}
Samuel Epstein.
\newblock All sampling methods produce outliers.
\newblock {\em IEEE Transactions on Information Theory}, 67(11):7568--7578,
  2021.

\bibitem[G\'21]{Gacs21}
Peter G\'{a}cs.
\newblock Lecture notes on descriptional complexity and randomness.
\newblock {\em CoRR}, abs/2105.04704, 2021.

\bibitem[Kol65]{Kolmogorov65}
A.~N. Kolmogorov.
\newblock {Three approaches to the quantitative definition of information.}
\newblock {\em {Problems in Information Transmission}}, 1:1--7, 1965.

\bibitem[Lev74]{Levin74}
L.~A. Levin.
\newblock {Laws of Information Conservation (Non-growth) and Aspects of the
  Foundations of Probability Theory}.
\newblock {\em Problemy Peredachi Informatsii}, 10(3):206--210, 1974.

\bibitem[Lev84]{Levin84}
L.~A. Levin.
\newblock {Randomness conservation inequalities; information and independence
  in mathematical theories}.
\newblock {\em {Information and Control}}, 61(1):15--37, 1984.

\bibitem[Lev13]{Levin13}
L.~A. Levin.
\newblock Forbidden information.
\newblock {\em J. ACM}, 60(2), 2013.

\bibitem[VV02]{VereshchaginVi02}
N.~Vereshchagin and P.~Vit{\'a}nyi.
\newblock Kolmogorov's structure functions and model selection.
\newblock {\em IEEE Transactions on Information Theory}, 50:3265--3290, 2002.

\bibitem[ZL70]{ZvonkinLe70}
A.~K. Zvonkin and L.~A. Levin.
\newblock The complexity of finite objects and the development of the concepts
  of information and randomness by means of the theory of algorithms.
\newblock {\em Russian Math. Surveys}, page~11, 1970.

\end{thebibliography}
\end{document}